\documentclass[runningheads]{llncs}
\usepackage[utf8]{inputenc}
\usepackage{enumitem}
\usepackage{amsmath}
\usepackage{graphicx}
\usepackage{subfigure}
\usepackage{multirow}
\usepackage{booktabs}
\usepackage{indentfirst}
\usepackage{makecell}
\usepackage{graphicx}
\usepackage{subfigure}
\usepackage{xcolor}
\usepackage{amssymb}

\newcommand{\ie}{\emph{i.e.,~}}
\newcommand{\etal}{\emph{et al.~}}

\begin{document}

\title{Two-Stream CNN with Loose Pair Training for Multi-modal AMD Categorization}
\titlerunning{Two-Stream CNN for Multi-modal AMD Categorization}
\newcommand*\samethanks[1][\value{footnote}]{\footnotemark[#1]}
\author{Weisen Wang\thanks{Equal Contribution}\textsuperscript{1,2,3}
\and Zhiyan Xu\samethanks\textsuperscript{4,5}
\and Weihong Yu\samethanks\textsuperscript{4,5}
\and Jianchun Zhao\textsuperscript{3} 
\and Jingyuan Yang\textsuperscript{4,5} 
\and Feng He\textsuperscript{4,5} 
\and Zhikun Yang\textsuperscript{4,5} 
\and Di Chen\textsuperscript{4,5} 
\and Dayong Ding\textsuperscript{3}
\and Youxin Chen\textsuperscript{4,5}
\and Xirong Li\thanks{\emph{Corresponding author: Xirong Li (xirong@ruc.edu.cn)}}\textsuperscript{1,2,3}}
\authorrunning{Weisen Wang, Zhiyan Xu, Weihong Yu et al.}
\institute{
\textsuperscript{1}MOE Key Lab of DEKE, Renmin University of China \\
\textsuperscript{2}AI \& Media Computing Lab, School of Information, Renmin University of China \\
\textsuperscript{3}Vistel AI Lab, Visionary Intelligence Ltd. \\
\textsuperscript{4}Key Lab of Ocular Fundus Disease, Chinese Academy of Medical Sciences\\
\textsuperscript{5}Department of Ophthalmology,
Peking Union Medical College Hospital}
\maketitle 

\begin{abstract}
This paper studies automated categorization of age-related macular degeneration (AMD) given a multi-modal input, which consists of a color fundus image and an optical coherence tomography (OCT) image from a specific eye. Previous work uses a traditional method, comprised of feature extraction and classifier training that cannot be optimized jointly. By contrast, we propose a two-stream convolutional neural network (CNN) that is end-to-end. The CNN's fusion layer is tailored to the need of fusing information from the fundus and OCT streams. For generating more multi-modal training instances, we introduce Loose Pair training, where a fundus image and an OCT image are paired based on class labels rather than eyes. Moreover, for a visual interpretation of how the individual modalities make contributions, we extend the class activation mapping technique to the multi-modal scenario. Experiments on a real-world dataset collected from an outpatient clinic justify the viability of our proposal for multi-modal AMD categorization. 
\end{abstract}


\section{Introduction} \label{sec:intro}
This paper targets at automated categorization of age-related macular degeneration (AMD). As a common macular disease among people over 50, AMD may cause blurred vision or even blindness if not treated in time \cite{Wan2014Global}. Depending on whether the retina contains choroidal neovascularization, AMD is classified into two subcategories, \ie \emph{dry AMD} (non-neovascular) and \emph{wet AMD} (neovascular) \cite{Ferris2013Clinical}. Due to different treatments, such a fine-grained classification is crucial. In the clinical practice, color fundus photography and optical coherence tomography (OCT) are used by an ophthalmologist to assess the condition of an eye. Not surprisingly, the lack of experienced ophthalmologists has driven the research towards automated AMD categorization based on either fundus images, OCT images or both. 

The majority of previous works are based on a single modality, let it be color fundus images capturing the posterior pole \cite{Burlina2016Detection,Burlina2017Comparing,Burlina2017Automated,Grassmann2018A} or OCT images \cite{Lee2017Deep,Karri2017Transfer,Treder2018Automated,Russakoff2019Deep,Kermany2018Identifying}. In \cite{Burlina2016Detection}, for instance,  Burlina \etal employ a deep convolutional neural network (CNN) pretrained on ImageNet to extract visual features from fundus images and then train a linear SVM classifier. As for OCT-based methods,  Lee \etal \cite{Lee2017Deep} train a VGG16 model to classify OCT images either as \emph{normal} or as \emph{AMD}. 
Since fundus images capture the state of the retinal plane, while OCT images reflect the longitudinal section of the retina, they describe distinct aspects of the retina and can thus be complementary to each other. While jointly exploiting the two modalities seems to be natural, this direction is largely unexplored. To the best of our knowledge,  Yoo \etal \cite{YooThe} make an initial attempt towards multi-modal AMD categorization. Given a pair of fundus and OCT images from a specific eye, the authors employ a VGG19 model pretrained on ImageNet to extract visual features from both images. The features are concatenated and used as input of a random forest classifier. Despite their encouraging result that the multi-modal method is better than its single-modal counterpart, some crucial questions remain open. 

Note that both the VGG19 features and the classifier, \ie random forest, used in \cite{YooThe} are suboptimal in the context of deep learning based visual categorization. The following questions arise. First, when the single-modal baseline is re-implemented using a state-of-the-art CNN, say ResNet \cite{He2016Deep}, in an end-to-end manner, is the multi-modal method by \cite{YooThe} still better? If the answer is negative, a follow-up question is can multi-modal AMD categorization be performed end-to-end as well? Training a deep network with multi-modal input is nontrivial because by definition, the number of paired multi-modal training instances is less than the number of single-modal training instances. Moreover, the method by \cite{YooThe} lacks the capability of  interpreting how the individual modalities contribute to the final prediction. 

\begin{figure}[tb!]
\centering
\includegraphics[width=\columnwidth]{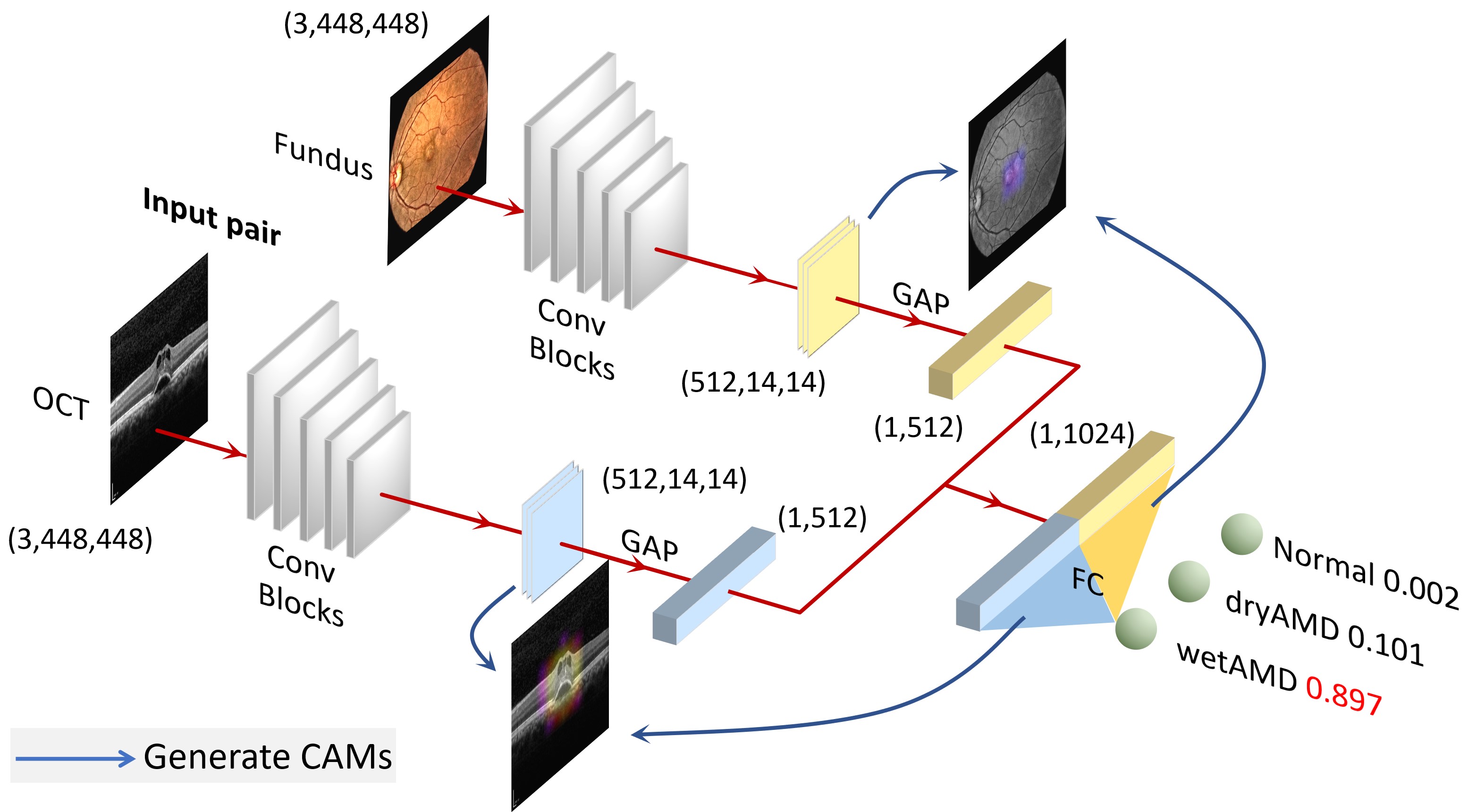}
\caption{\textbf{A conceptual diagram of the proposed two-stream CNN for multi-modal AMD categorization}. The network consists of two symmetric branches, one for processing fundus images while the other for processing OCT images. Given a pair of fundus and OCT images taken from a specific eye, the proposed network makes a three-class prediction concerning the probability of the eye being \emph{normal}, \emph{dryAMD} and \emph{wetAMD}, respectively. Moreover, we adopt  class activation mapping (CAM) \cite{Zhou2015Learning} to visually interpret how the multi-modal input contributes to the prediction.}
\label{fig:framework}
\end{figure}

Towards answering the above questions, we make contributions as follows.
\begin{itemize}
\item We propose a two-stream CNN specifically designed for multi-modal AMD categorization, see Fig. \ref{fig:framework}. Two-stream CNNs have been actively investigated in the context of video action recognition \cite{feichtenhofer2016convolutional}. However, the fusion layer needs to be re-considered for the new task, not only for effectively combining the information from fundus and OCT images but also for visually interpreting  their contributions. 
\item To attack the inadequacy of multi-modal training instances, we introduce Loose Pair Training, a simple sampling strategy that effectively increases the number of training instances.
\item Experiments on real-world data collected from an outpatient clinic show the viability of the proposed method.  The new method outperforms the state-of-the-art \cite{YooThe} with a large margin, \ie 0.971 \emph{versus} 0.826 in terms of overall accuracy, for multi-modal AMD categorization. 
\end{itemize}

\section{Our Method} \label{sec:method}
Given a color fundus image $I_f$ and an OCT image $I_o$ taken from a specific eye, we aim to build a multi-modal CNN (MM-CNN) that takes the paired input and categorizes the eye's condition to a specific class $c$:
\begin{equation} \label{eq:general}
c \leftarrow \mbox{MM-CNN}(\{I_f, I_o\}),
\end{equation}
with $c \in \{normal, dryAMD, wetAMD\}$. 

\subsection{Multi-modal CNN} \label{ssec:network}

\textbf{Network architecture}. To handle the multi-modal input, we design a two-stream network as illustrated in Fig. \ref{fig:framework}. It consists of two symmetric branches, one for processing the fundus image $I_f$ and the other for processing the OCT image $I_o$. Note that such an architecture resembles to some extent the two-stream network widely used for video action recognition \cite{feichtenhofer2016convolutional}. The major difference is at which layer multi-modal fusion is performed. Feature maps generated by intermediate layers of a CNN preserves, to some extent, the spatial information of an input image. As different streams of video data are spatially correlated, the state-of-the-art for video action recognition performs fusion by combining feature maps from the individual streams \cite{feichtenhofer2016convolutional}. By contrast, as $I_f$ and $I_o$ are not spatially correlated, we opt to perform the fusion after the global average pooling (GAP) layer, which removes the spatial information by averaging each feature map into a single value.

For each branch, we use convolutional blocks of ResNet-18 \cite{He2016Deep}. In principle, any other state-of-the-art CNN can be used here. We choose ResNet-18 as it has fewer parameters and thus requires less training data. Also, this CNN is shown to be effective for other fundus image analysis tasks \cite{mmm2019-left-right-eye}. 
For an OCT image, we convert each of its pixels from grayscale to RGB by duplicating the intensity for each RGB component. As such, the same architecture and  initialization are applied to both branches.

Let $\mathbf{F}_f=\{F_{f,1},\ldots, F_{f,512}\}$ be an array of $m \times m$ feature maps generated by the ResNet-18 module in the fundus branch. The value of $m$ depends on the size of the input, which is $14$ for an input size of $448 \times 448$. Given a specific feature map $F_{f,i}$, the value of a specific position $(x,y)$ is acquired as $F_{f,i}(x,y)$. In a similar vein, we define the feature maps for the OCT branch as $\mathbf{F}_o=\{F_{o,1},\ldots, F_{o,512}\}$. 

Our fusion layer is implemented by first feeding separately $\mathbf{F}_f$ and $\mathbf{F}_o$ into a GAP layer to obtain two $1\times 512$ vectors, denoted as $(\bar{F}_{f,1}, \ldots, \bar{F}_{f,512})$ and $(\bar{F}_{o,1}, \ldots, \bar{F}_{o,512})$, respectively. The two vectors are then concatenated to form a $1 \times 1024$ vector which contains information from the two modalities. For classification, the combined vector is fed into a fully connected (FC) layer to produce a score for a specific class $c$, denoted as $s^c$, 
\begin{equation} \label{eq:raw-score}
s^c = \sum^{512}_{i=1} w^c_{f,i} \cdot \bar{F}_{f,i} + \sum^{512}_{i=1} w^c_{o,i} \cdot \bar{F}_{o,i},
\end{equation}
where $\{w^c_{f,1},\ldots,w^c_{f,512}\}$ and $\{w^c_{o,1},\ldots,w^c_{o,512}\}$ are class-dependent weights parameterizing the FC layer. Classification as expressed in Eq. \ref{eq:general} is achieved by selecting the class with the maximum score.

\textbf{Multi-modal class activation mapping for visual interpretation}. As Eq. \ref{eq:raw-score} shows, the classification score $s^c$ for a given class $c$ is additively contributed by both modalities. For a more intuitive interpretation, we leverage class activation mapping (CAM) \cite{Zhou2015Learning}, which reveals the (implicit) attention of a CNN on an input image. We compute the multi-modal version of CAMs as 
\begin{equation}\label{eq:cam}
\left\{
\begin{array}{lr}
CAM^{c}_{f}(x,y)=\sum^{512}_{i=1} w^{c}_{f,i} \cdot F_{f,i}(x,y),\\
\\
CAM^{c}_{o}(x,y)=\sum^{512}_{i=1} w^{c}_{o,i} \cdot F_{o,i}(x,y).
\end{array}
\right.
\end{equation}

\begin{figure}[tb!]
\centering
\subfigure[Multi-modal input]{
\label{fig:instance:a}
\includegraphics[width=0.28\columnwidth]{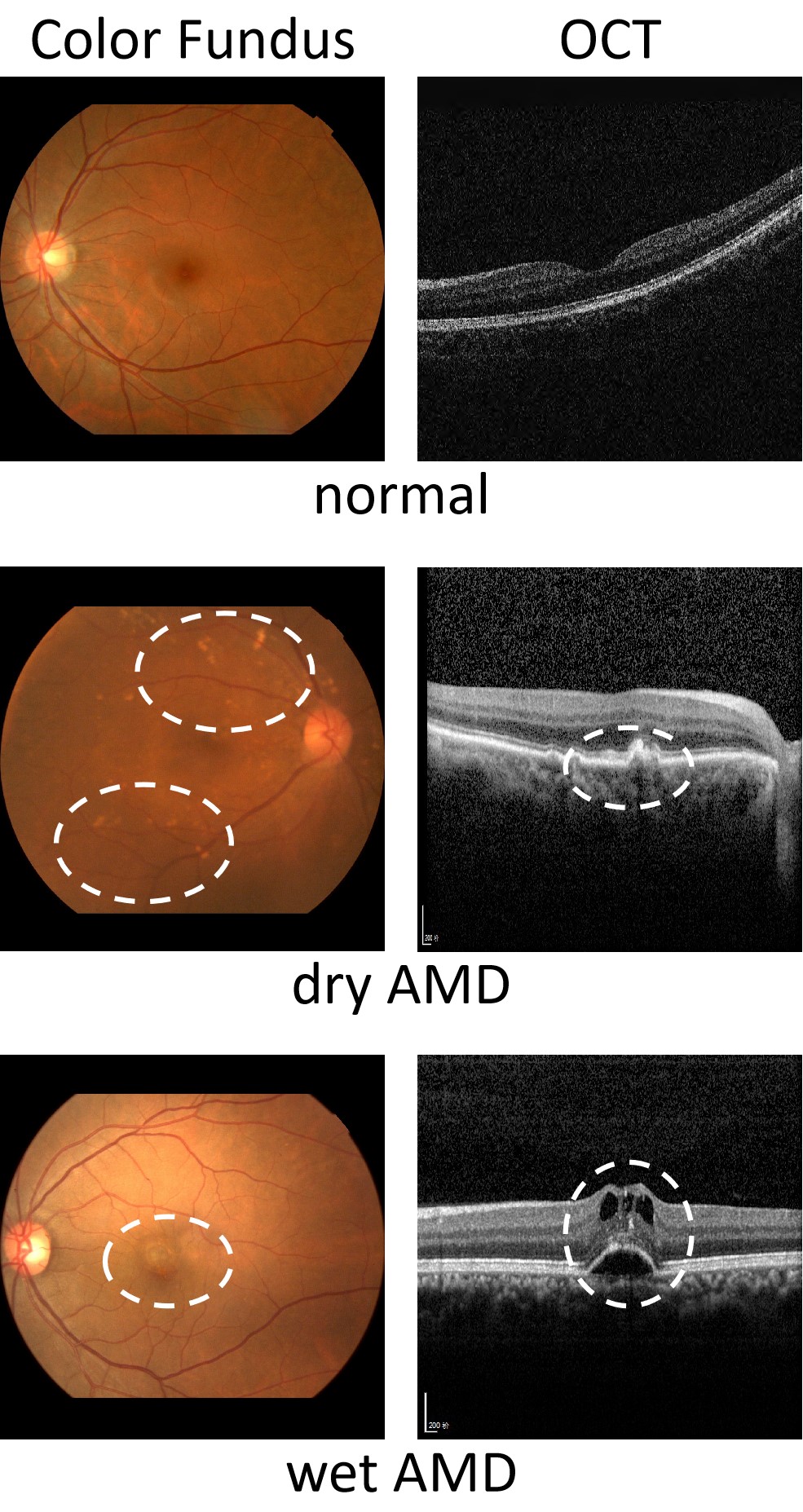}}
\subfigure[Single-modal CAMs]{
\label{fig:instance:b}
\includegraphics[width=0.28\columnwidth]{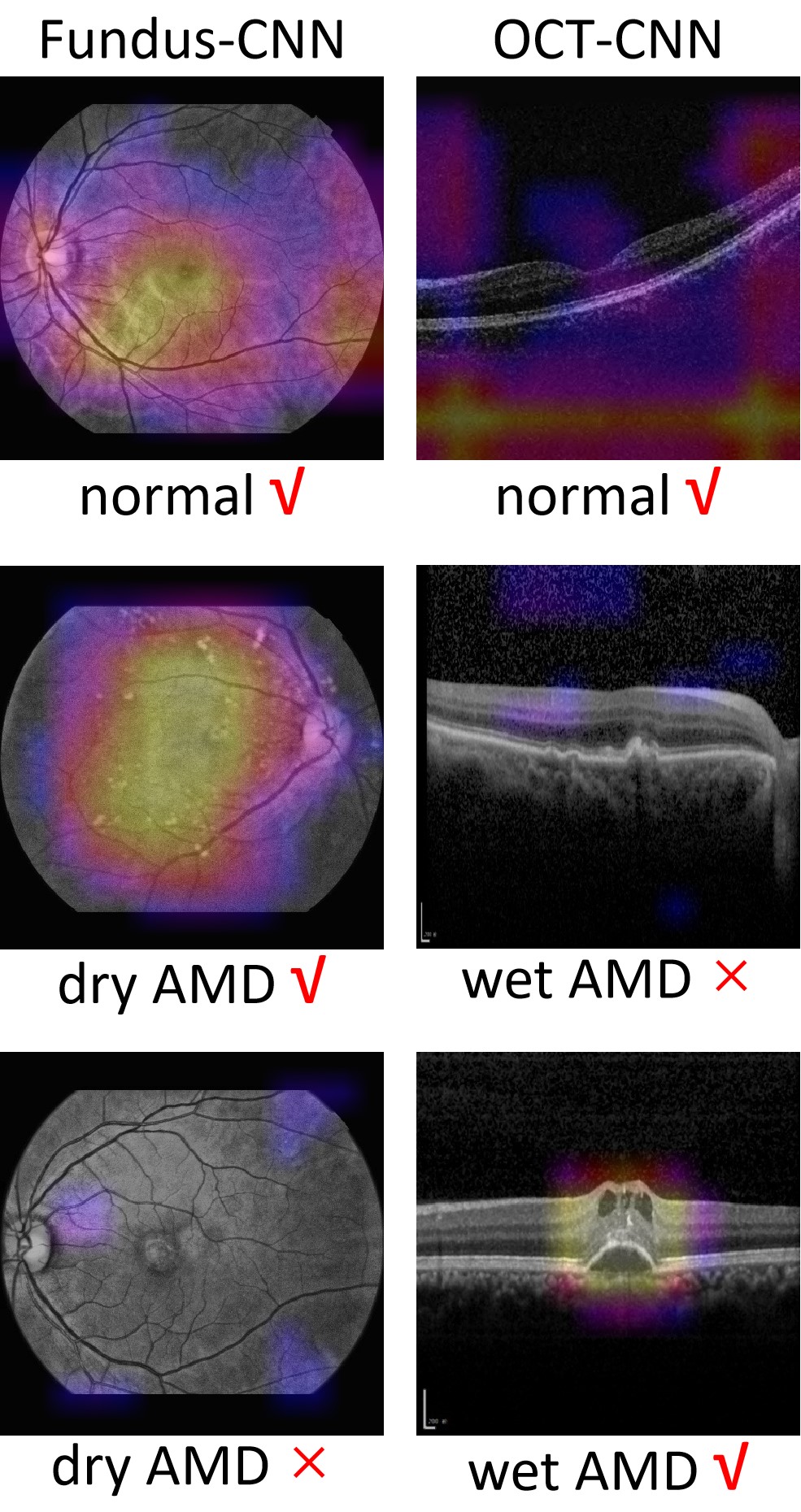}}
\subfigure[Multi-modal CAMs]{
\label{fig:instance:c}
\includegraphics[width=0.28\columnwidth]{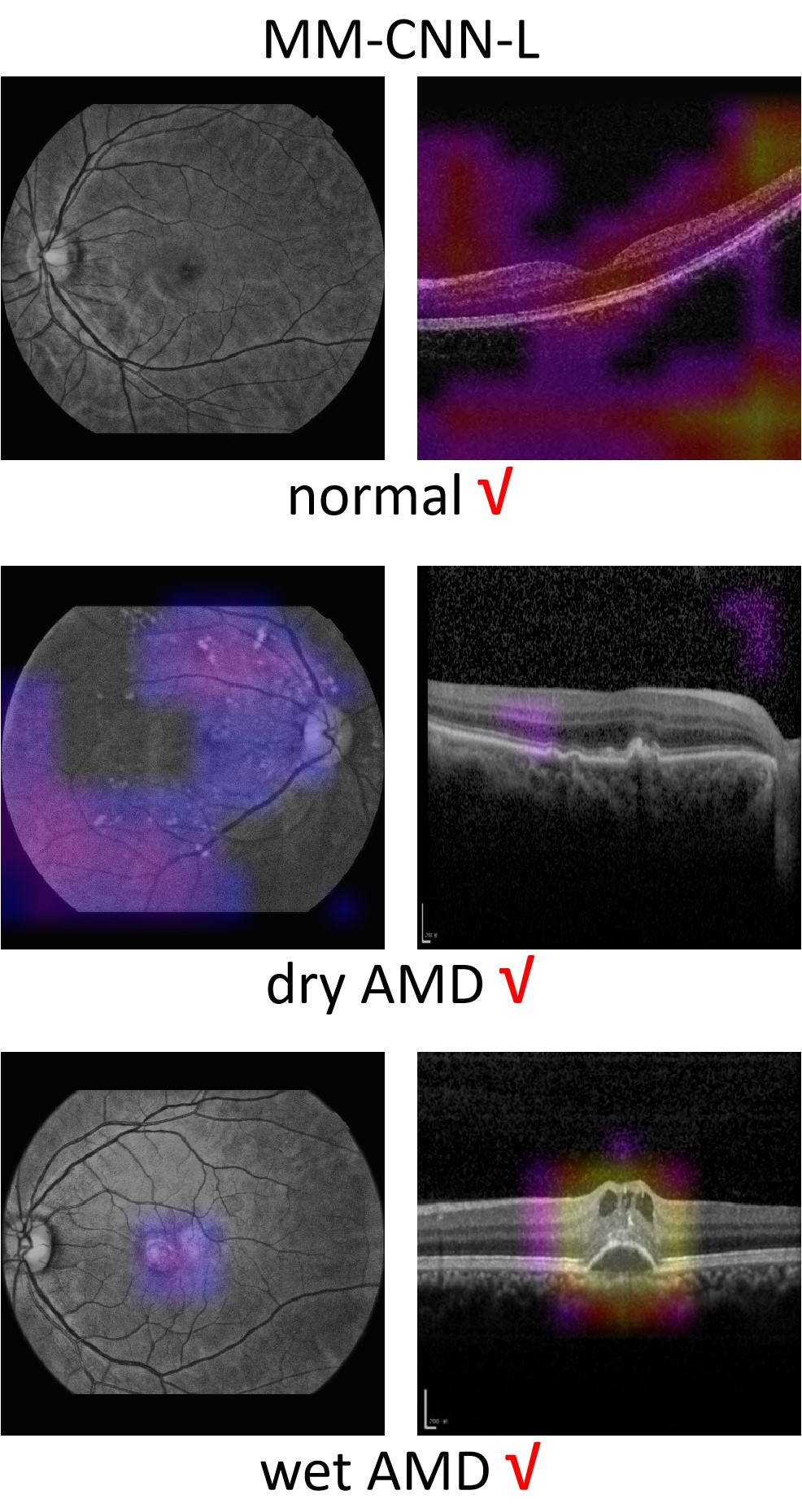}}
\caption{\textbf{CAM-based visualization of single-modal / multi-modal CNNs}. White ellipses in (a) indicate regions related to a specific AMD class. Brighter areas in (b) and (c) indicate higher activations. MM-CNN-L is the proposed multi-modal CNN with loose pair training. Note that the color fundus images are shown in gray for better visualizing the heat maps. Best viewed in digital format.}
\label{fig:instance}
\end{figure}

Note that $ \bar{F}_{f,i} = \sum_{x,y} F_{f,i}(x,y) $ and $ \bar{F}_{f,o} = \sum_{x,y} F_{f,o}(x,y) $. Putting Eq. \ref{eq:raw-score} and \ref{eq:cam} together, $s^c$ can be rewritten as 
\begin{equation} \label{eq:sc-cam}
s^c = \sum_{x,y} CAM^{c}_{f}(x,y) + \sum_{x,y} CAM^{c}_{o}(x,y).
\end{equation}
According to Eq. \ref{eq:sc-cam}, $CAM^{c}_{f}(x,y)$ and $CAM^{c}_{o}(x,y)$ indicate the contribution of a specific position of the fundus and OCT images, respectively. Consequently, the contribution of each modality can be visualized by overlaying with the corresponding up-sampled CAM, see Fig. \ref{fig:instance}.

\subsection{Network Training} \label{ssec:training}

A conventional way to construct a multi-modal training instance is to strictly select a fundus image and an OCT image from the same eye, which we term strict pairing. By contrast, we construct instances based on labels instead of eyes. That is, a fundus image is allowed to be paired with an OCT image if their labels are identical. We coin this sampling strategy \emph{Loose Pairing}. Such a strategy expands the size of the training set quadratically. Note that loose pairing is applied only on the training data.

All fundus and OCT images are resized to $448 \times 448$. As the input of the pretrained ResNet-18 model is $224 \times 224$, we adjust the kernel size of the GAP layer from $7 \times 7$ to $14 \times 14$. Following \cite{Jintasuttisak2014Color}, we enhance fundus images by contrast-limited adaptive histogram equalization. Meanwhile, median filtering is applied on OCT images for noise reduction.  For image-level data augmentation, random rotation, crop, flip and random changes in brightness, saturation and contrast are performed on training images.

Our deep models are implemented in the PyTorch (version 1.0.0) framework. ResNet-18 was pretrained on ImageNet. We use cross-entropy, a common loss function for multi-class classification. SGD with momentum of 0.9 and weight decay of 1e-4 is used as the optimizer. Each convolution layer is followed by batch normalization. No dropout is used. The model that obtaining the best validation performance is selected.

\section{Evaluations} \label{sec:eval}
\subsection{Experimental Setup}

\textbf{Dataset for multi-modal AMD categorization}. We collect 1,059 color fundus images from 1,059 distinct eyes at the outpatient clinic of the Department of Ophthalmology, Peking Union Medical College Hospital. That is, one fundus image per eye. For 781 eyes, they are associated with one to five OCT images, which are central B-scans manually selected by technicians. The fundus images were acquired from a Topcon fundus camera, while OCT images came from a Topcon OCT camera and a Heidelberg OCT camera. For each eye, two ophthalmologists jointly classify its condition as \emph{normal}, \emph{dryAMD} or \emph{wetAMD}, by examining the corresponding fundus image plus OCT, fluorescein angiography (FA) or indocyanine green angiography (ICGA) images, if applicable. Fundus and OCT images associated with a specific eye are assigned with the same class. 

In order to build a multi-modal test set, per class we select 20 eyes at random from the eyes that have both fundus and OCT images available. Such a setting  allows us to justify the effectiveness of multi-modal input against its single-modal counterpart. Moreover, it enables a head-to-head comparison between the two single modalities, \ie fundus versus OCT. In a similar vein, we construct a multi-modal validation set from the remaining data for model selection. All the rest is used for training. Table \ref{tab:data} shows data statistics.

\begin{table}[tb!]
\renewcommand{\arraystretch}{1.2}
\centering
\caption{\textbf{Dataset used in our experiments}. Data split is made based on eyes. In parentheses are number of eyes per class in each split.}\label{tab:data}
\scalebox{0.9}{
\begin{tabular}{ll rrrr rrrr rrr}
\toprule
   
                      && \multicolumn{3}{c}{\textbf{Training images}} && \multicolumn{3}{c}{\textbf{Validation images}}  && \multicolumn{3}{c}{\textbf{Test images}}  \\

                      \cmidrule{3-5} \cmidrule{7-9} \cmidrule{11-13}

 \textbf{Class}  && \textit{Fundus} && \textit{OCT} && \textit{Fundus} && \textit{OCT}    && \textit{Fundus} && \textit{OCT} \\
\midrule
\emph{normal}       && 155 (155) && 156 (155)  && 20 (20) && 20 (20)  && 20 (20) && 20 (20) \\
\emph{dryAMD}      && 67 (~67)   && 33 (~22)    && 20 (20) && 35 (20)  && 20 (20) && 38 (20) \\
\emph{wetAMD}      && 717 (717) && 821 (484)  && 20 (20) && 42 (20)  && 20 (20) && 46 (20) \\
\bottomrule
\end{tabular}
}
\end{table}

\textbf{Performance metrics}. Per class we report three metrics, \ie sensitivity, specificity and F1 score defined as the harmonic mean between sensitivity and specificity. For an overall comparison, the average F1 score over the three classes is used. In addition, we report accuracy, computed as the ratio of correctly classified instances (which are fundus or OCT images for single-modal CNNs and fundus-OCT pairs for MM-CNNs). 

\subsection{Experiment 1. Multi-modal \emph{versus} Single-modal} \label{ssec:exp-mm-vs-single}

\textbf{Single-modal baselines}. For single-modal models, we train two ResNet-18 on the fundus images and the OCT images, respectively. For the ease of reference we term the two models Fundus-CNN and OCT-CNN.

\textbf{Results}. As Table \ref{tab:results} shows, OCT-CNN is on par with MM-CNN-S, which is trained on the strict pairs. The result suggests that training an effective multi-modal model requires more training data. The proposed loose pair training strategy is effective, resulting in MM-CNN-L that presenting the best performance. 

Comparing the two single-modal models, OCT-CNN is better than Fundus-CNN (0.942 versus 0.879 in terms of the overall F1). Confusion matrices are provided in the supplementary material. 
While the two single-modal CNNs recognize the normal class with ease, they tend to misclassify dryAMD as wetAMD. Such mistakes are reduced by MM-CNN-L. The above results justify the advantage of multi-modal models for AMD categorization. 

\begin{table}[tb!]
\renewcommand\arraystretch{1.2}
\centering
\caption{\textbf{Performance of different models on the test set}. MM-CNN-L, which is the proposed multi-modal CNN with loose pair training, performs the best.}
\label{tab:results}
\scalebox{0.85}{
\begin{tabular}{@{}ll rrr  rrr rrr rr@{}}
\toprule
      
    && \multicolumn{3}{c}{\textbf{Normal}}  & \multicolumn{3}{c}{\textbf{dryAMD}}  & \multicolumn{3}{c}{\textbf{wetAMD}}  & \multicolumn{2}{c}{\textbf{Overall}}\\
   \cmidrule{3-5} \cmidrule(l){6-8} \cmidrule(l){9-11} \cmidrule(l){12-13}
\textbf{Model} && \textit{Sen.}  & \textit{Spe.}  & \textit{F1}  & \textit{Sen.}  & \textit{Spe.}  & \textit{F1} & \textit{Sen.}  & \textit{Spe.}  & \textit{F1}  & \textit{F1}  & \textit{Accuracy}\\
\midrule
 
\textbf{Single-modal:} & \\
Fundus-CNN    && 1.000 & 0.975 & 0.975 & 0.700 & 0.975 & 0.800 & 0.950 & 0.875 & 0.863 & 0.879 & 0.883\\
OCT-CNN       && 1.000 & 1.000 & \textbf{1.000} & 0.815 & 1.000 & 0.898 & 1.000 & 0.879 & 0.929 & 0.942 & 0.932\\ [2pt]

\textbf{Multi-modal:} & \\
Yoo \etal  \cite{YooThe} && 1.000 & 0.976 & 0.952 & 0.552 & 1.000 & 0.711 & 0.978 & 0.724 & 0.841 & 0.835 & 0.826\\
Yoo \etal-L        && 1.000 & 0.988 & 0.975 & 0.763 & 0.954 & 0.828 & 0.913 & 0.844 & 0.866 & 0.890 & 0.875\\ [2pt]
\emph{MM-CNN-S}   && 1.000 & 1.000 & \textbf{1.000} & 0.842 & 0.984 & 0.901 & 0.978 & 0.896 & 0.927 & 0.943 & 0.932\\
\emph{MM-CNN-L}  && 1.000 & 1.000 & \textbf{1.000} & 0.921 & 1.000 & \textbf{0.958} & 1.000 & 0.948 & \textbf{0.968} & \textbf{0.975} & \textbf{0.971}\\

\bottomrule
\end{tabular}
}
\end{table}

\subsection{Experiments 2. Comparison with the State-of-the-art} \label{ssec:exp-sota}

\textbf{Multi-modal baselines}. As aforementioned, the only existing work on multi-modal AMD categorization is by Yoo \etal \cite{YooThe}, where the authors employ a VGGNet pretrained on ImageNet to extract visual features from fundus and OCT images and then train a random forest classifier on strictly matched pairs.  We therefore consider that work as our multi-modal baseline. As their data is not fully available, we replicate their method and evaluate on our test set. For a fair comparison, we substitute ResNet-18 for VGGNet. Moreover, we investigate if the proposed loose pair strategy is also beneficial for the baseline. So we train another random forest with loose pairs. We term this variant Yoo \etal -L. 

\textbf{Results}. As Table \ref{tab:results} shows, MM-CNN-L outperforms the baseline with a large margin (0.975 \emph{versus} 0.835 in terms of overall F1). The two single-modal baselines outperform Yoo \etal \cite{YooThe}. These results justify the necessity of end-to-end learning. The loose pair training strategy is found to be useful for the baseline also, improving its overall F1 from 0.835 to 0.890.

\section{Conclusions} \label{sec:conc}
Multi-modal AMD categorization experiments on a clinical dataset allow us to answer the questions asked in Section \ref{sec:intro} as follows. When end-to-end trained, a single-modal CNN, in particular OCT-CNN, is a nontrivial baseline to beat. Multi-modal CNN recognizes dry AMD and wet AMD at a higher accuracy. This advantage is  obtained by the proposed two-stream CNN with loose pair training. 

\medskip

\textbf{Acknowledgments}. This work was supported by NSFC (No. 61672523), the Fundamental Research Funds for the Central Universities and the Research Funds of Renmin University of China (No. 18XNLG19), and CAMS Initiative for Innovative Medicine (No. 2018-I2M-AI-001).

\end{document}